\documentclass[10pt,conference]{IEEEtran}
\IEEEoverridecommandlockouts
\usepackage{cite}
\usepackage{float}
\usepackage{amsmath,amssymb,amsfonts}
\usepackage{mathtools}
\usepackage{algorithmic}
\usepackage{graphicx}
\usepackage{subcaption}
\usepackage{textcomp}
\usepackage{xcolor}
\usepackage{tabularx,ragged2e,booktabs}
\usepackage{multirow}
\usepackage{rotating}
\usepackage[utf8]{inputenc}
\usepackage{dblfloatfix}
\usepackage{ulem}
\usepackage[hang,flushmargin]{footmisc}
\usepackage[font=small,labelfont=bf]{caption}
 
\def\BibTeX{{\rm B\kern-.05em{\sc i\kern-.025em b}\kern-.08em
    T\kern-.1667em\lower.7ex\hbox{E}\kern-.125emX}}

\begin{document}

\title{On-line Anomaly Detection and Qualification of Random Bit Streams}

\author{
    \IEEEauthorblockN{Cesare Gerolimetto Fabrello\IEEEauthorrefmark{1}, Valeria Rossi\IEEEauthorrefmark{1}, Kamil Witek\IEEEauthorrefmark{1}\IEEEauthorrefmark{2}, Alberto Trombetta\IEEEauthorrefmark{1}, and Massimo Caccia\IEEEauthorrefmark{1}}
    \IEEEauthorblockA{\IEEEauthorrefmark{1}Università degli Studi dell'Insubria}
    \IEEEauthorblockA{\IEEEauthorrefmark{2}AGH University of Krakow}
}

\maketitle
    
\begin{abstract}
Generating random bit streams is required in various applications, most notably cyber-security. Ensuring high-quality and robust randomness is crucial to mitigate risks associated with predictability and system compromise. True random numbers provide the highest unpredictability levels. However, potential biases in the processes exploited for the random number generation must be carefully monitored. This paper reports the implementation and characterization of an on-line procedure for the detection of anomalies in a true random bit stream. It is based on the NIST Adaptive Proportion and Repetition Count tests, complemented by statistical analysis relying on the Monobit and RUNS. The procedure is firmware implemented and performed simultaneously with the bit stream generation, and providing as well an estimate of the entropy of the source. The experimental validation of the approach is performed upon the bit streams generated by a quantum, silicon-based entropy source.
\end{abstract}


\thispagestyle{plain}
\pagestyle{plain}
\pagenumbering{gobble}

\section{Introduction}
\label{sec:intro}
The use of massive amounts of random numbers is a critical issue in security-related techniques and tools for protecting and sharing data in large, distributed environments \cite{gennaro}, as well as their deployment in statistical and numerical simulations \cite{prng-need3}.

The need for high-quality random numbers has led to the development of True Random Number Generators (TRNGs) and Pseudo Random Number Generators (PRNGs). TRNGs exploit unpredictability of classes of natural phenomena, due to stochastic or quantum effects.
Generators employing the latter are referred to as Quantum Random Number Generators (QRNGs). PRNGs, or Deterministic Random Bits Generators (DRBGs), use algorithms to mimic TRNGs but are inherently limited in unpredictability.

Proving randomness poses challenges regarding diagnostic statistical methods and their implementation. Several statistical tests have been devised, notably the \textit{Statistical Test Suite} by the National Institute of Standards and Technology (NIST) \cite{nist800-22} and \textit{TestU01} \cite{testu01}. While these tests offer an extended and detailed assessment of entropy, they require a substantial volume of bits and are computationally demanding. Therefore, they are not suitable for an on-line quality estimation of a stream performed simultaneously with the generation. To address this issue, the NIST has established guidelines for performing continuous health tests \cite{nist800-90b}.

Regardless of the random bit-stream generation mechanism of random bits, the significant impact of randomness quality in security applications emerged from several use cases. A well-known example that surfaced in 2008 \cite{debian} concerns a critical vulnerability within the Debian Linux release of OpenSSL, resulting in low entropy during cryptographic key generation. Despite it being discovered and promptly fixed, the response was sluggish, and certificate authorities persisted in issuing authentications with weak keys even after the vulnerability was disclosed \cite{debian2}. 
Recent work \cite{iotrng} reports that errors in cryptographic key generation in IoT devices due to a slow rate of random bit harvesting went unnoticed, leading to significant entropy loss affecting the security of billions of devices.
Almost surely, such vulnerabilities could have been promptly detected by implementing on-line randomness quality estimation methods. This need is actually recognized by NIST, which prescribes in the DRBG \cite{nist800-90a} and TRNG \cite{nist800-90c} procedures the implementation of "health tests", namely quality assessments of possibly limited sensitivity, but rapid execution.

\underline{\textbf{Our contributions:}} In this paper, we present an anomaly detection procedure using outputs from NIST health tests, namely the Repetition Count Test (RCT) and Adaptive Proportion Test (APT), complemented by Monobit and RUNS statistics. We enhance the analysis of source entropy by statistically interpreting Monobit and RUNS results by measuring the shift of the average of a sample from the expected value for an unbiased sequence of bits, and proposing a new method to estimate a lower bound on source entropy using RCT and APT failure frequencies. These procedures are implemented in FPGA firmware, ensuring on-line execution without affecting the bit generation rate.

The paper is organized as follows: in Section \ref{sec:stat_tests} a short introduction to the statistical tests is presented; Section \ref{sec:methods} describes the experimental procedure and the results obtained using a Silicon-based QRNG are reported; finally, conclusions and outlook are drawn in Section \ref{sec:conc}.

\section{Standard On-line Statistical Tests}
\label{sec:stat_tests}
The anomaly detection procedure, presented in Section \ref{sec:methods}, is based on a set of four well-known and recognized tests, operating on single sets of bits or symbols. The Monobit and RUNs tests were selected from the numerous tests within the NIST test suite based on their simplicity, essential for a firmware implementation, and the complementary information they offer. Specifically, the Monobit test assesses the asymmetry of bit distribution within a bit-string, whereas the RUNs test quantifies the occurrence of bit-flips. Due to their minimal correlation, these tests maximize the information inferred from the data. These two tests find their natural generalization to symbols, respectively in the Adaptive Proportion and in the Repetition Count tests, which are defined by NIST in \cite{nist800-90b} and used to asses the eligibility of a sequence to seed a DRBG.

\subsection{Symmetry tests}

\subsubsection{Monobit}
The Monobit test \cite{nist800-22} asserts the asymmetry between zeros and ones in a bit sequence. Given a set of $n$ bits, the stochastic variable is defined as 
\begin{equation}
S_{n} = \sum_{i=1}^{n}x_{i} = 2n_1 - n\label{eqMono}
\end{equation}
where $x_{i} = 2 \epsilon_{i} - 1$,  $\epsilon_{i}$ is the bit state and $n_1$ is the number of 1s in the bit sequence. Provided that the bit values are independent and identically distributed (i.i.d.), the number of bits set to 1 follows the binomial probability density function: 

\begin{equation}
B(n_{1}, n, p) = \frac{n!}{n_{1}!(n-n_{1})!}p^{n_{1}}(1-p)^{n-n_{1}},\label{eqBin}
\end{equation}
with $p$ being the probability of generating 1. If both values are equally probable, then $\overline{S_n} = 0 $, with standard deviation $\sigma_{S_{n}} = \sqrt{n}$. The Monobit test fails whenever a sequence has a value $S_n$ which exceeds an alarm level $k\sqrt{n}$, where k is defined according to the sensitivity and false alarm rate set by the user.

\subsubsection{Adaptive Proportion Test}
The APT expands upon the binary checks performed by the Monobit test to include an alphabet of $m$ symbols. As defined in the NIST documentation \cite{nist800-90b}, the stream is divided into sequences of length $n$. The frequency of the occurrence of the first symbol is checked against the hypothesis of a binomial distribution with $p = 1/m$.

Specifications for the sequence length to be considered are contingent on the alphabet used and are provided by NIST. For binary sequences the recommended length is $n=1024$, while for non-binary sequences it is suggested to use $n=512$. 
Within each sequence, the occurrences of the first symbol are counted, and compared to a cut-off threshold corresponding to a defined failure probability. The cut-off threshold can be calculated relying on the Binomial Cumulative Distribution Function, assuming an equal probability for each of the $m$ symbols to be generated.

\subsection{Runs Tests}
\subsubsection{RUNS}
\begin{figure}[tp]
\centerline{\includegraphics[width=0.5\textwidth]{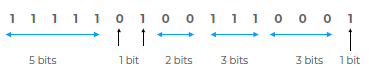}}
\caption{The RUNS test counts the number of sequences of consecutive identical bits in a bit-stream. In this figure, a sequence of $n=16$ bits containing a total of 7 runs is shown.}
\label{runsdefinition}
\end{figure}

This test counts the number of series of consecutive identical bits in a sequence of specified length, as shown in Fig. \ref{runsdefinition}. It is worth noting that this quantity can be measured by the number of bit flips plus one. 
\\
This test complements the Monobit: for a given value of $n_1$, it measures the expected number of bit flips in the sequence in the hypothesis of "fair coin". The underlying probability distribution is the probability of having a runs of a given length conditioned on the number of ones in the sequence and $p = 1/2$. By defining \[\pi = \frac{n_1}{n}\] as the fraction of bits set to 1, the average number of runs and the variance are given by \cite{distfree-tests} 
\begin{align}
\begin{split}
    \overline{R} &= 2n\pi (1-\pi)+1, \\
    \sigma^2_R &= \frac{n}{n-1}\times[2\pi (1-\pi)(2n\pi(1-\pi)-1)].
\end{split}
\end{align}
Ultimately, the purpose of this test is to determine whether the oscillation between sub-sequences of identical bits is too fast or too slow from the expectations \cite{runsdef}.
As for the Monobit, whenever a sequence exceeds a threshold value $\overline{R}\pm k \sigma_R$ the test is considered failed.

\subsubsection{Repetition Count Test}
The RCT also relies on the concept of runs generalizing the test to sequences of symbols. However, instead of counting the number of symbol changes, it focuses on the length of consecutive identical instances.

The probability $\alpha$ of having at least $C$ consecutive equal symbols, can be written as: 
\begin{align}
\begin{split}\label{eqProbabilityC}
\alpha = P(k\geq C)= \sum_{i=1}^m (1-p_i) p_i p_i^{C-1} = \sum_{i=1}^m (1-p_i) p_i^{C}
\end{split}
\end{align}
where $k$ is the length of the runs, $p_i$ stands for the probability associated to the $i$-th symbol of the alphabet and $(1-p_i)$ represents the probability of the symbol generated before the current one to be different, resetting the counter of the length of the run.
Assuming $p_1 \geq p_2 \geq .... \geq p_m$, \eqref{eqProbabilityC} can be given an upper bound by:
\begin{equation}
 \alpha = P(k\geq C)\leq (m-1)\cdot p_1^{C}=(m-1)(2^{-H})^C, \label{eqAlpha}
\end{equation}
where $H$ is the min-entropy defined as $H=-\log_2(\max\{p_i\})$ \cite{nist800-90b}. Once the value of $\alpha$ is defined, the cut-off threshold $C$ can then be computed as:
\begin{equation}
C = \Bigg\lceil \frac{1}{H}\Bigg[log_{2}(m-1)-log_{2}(\alpha)\Bigg] \Bigg\rceil.\label{eqCRCT}
\end{equation}
It is worth noting that equation \eqref{eqCRCT} differs from the original NIST prescription as a consequence of the implementation, which stands on the assumption that the symbols are being analyzed sequentially. Consequently, it embodies the viewpoint of the ``observer symbol", which must differ from the preceding one and match the subsequent $C-1$ symbols.\\
NIST recommends choosing the $\alpha$ parameter between $2^{-20}$ and $2^{-40}$, equivalent to  respectively a $5\sigma$ and a $7\sigma$ confidence level presuming a Gaussian distribution of the measured quantities. In this work, a value of $\alpha = 2^{-20}$ is chosen.

\begin{figure}[t]
\centerline{\includegraphics[width=0.5\textwidth]{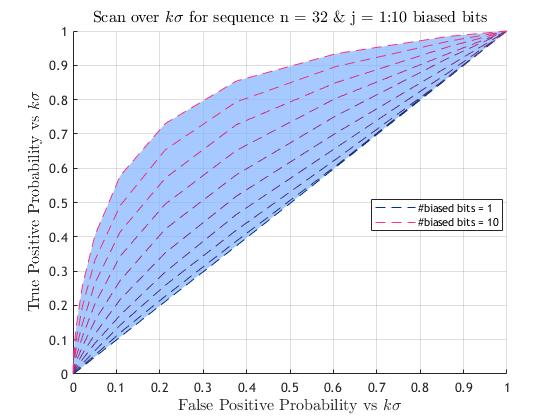}}
\caption{Sensitivity scan of the Monobit test across different confidence levels for a sequence of length $n=32$ with a number $j$ of biased bits ranging from 1 to 10.}
\label{false_vs_true_pos_k_scan}
\end{figure}

\section{Anomaly Detection Procedure and Experimental Results}
\label{sec:methods}

The anomaly detection procedures were applied to bit streams generated by a device extracting entropy through the analysis of the time series of self-amplified endogenous pulses due to stochastically generated charge carriers in an array of p-n junctions operated beyond the breakdown voltage, actually mimicking pulses originated by radioactive decays \cite{rap!}. Pulses are seeded by charge carriers crossing potential barriers, a well-modeled quantum phenomenon (see for instance  \cite{sipm1}, \cite{sipm2} and \cite{sipm3}), entering a high electric field region where they produce an avalanche by impact ionization, in quenched Geiger-M\"uller regime. 

This Silicon-based QRNG generates sequences of four bits analyzing the inter-arrival time of a series of nine pulses, resulting in an alphabet of $2^4$ possible symbols for each cycle, over which the analysis of the RCT and APT is performed, while the Monobit and RUNs analyze the raw bit-stream.
It is worth noting that correlations during the generation process, dead-times of the QRNG during time stamping of the pulses, and external factors, like thermal runaways, may introduce time-dependent anomalies, making the implementation of on-line health tests a relevant diagnostic tool.

The tests are implemented in FPGA to enable parallel execution of generation and randomness assessment, minimizing latency and maintaining the generation rate.
Furthermore, they check for biases in the data and assess the quality of the stream, acting on a series of sequences of bits of user-defined length.

\begin{figure}[t]
\centerline{\includegraphics[width=0.5\textwidth]{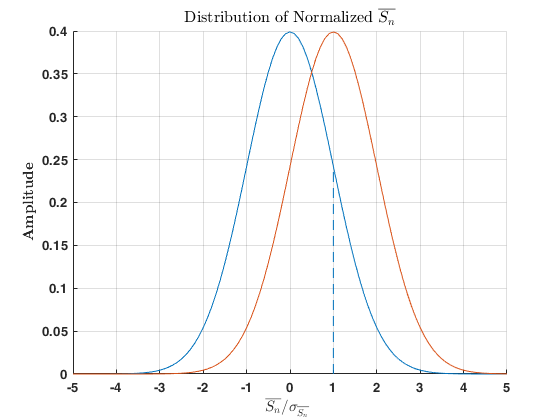}}
\caption{Exemplary distributions of the normalized $\overline{S_n}$ value for an unbiased (in blue) and biased (in red) bit stream. Two metrics were considered to distinguish the two: a shift in the mean value of the distribution and the number of events in the tails over a fixed threshold, here illustrated by the dashed line.}
\label{figZShift}
\end{figure}

\subsection{Monobit}

\begin{figure*}[tph]
\centerline{\includegraphics[width=1\textwidth]{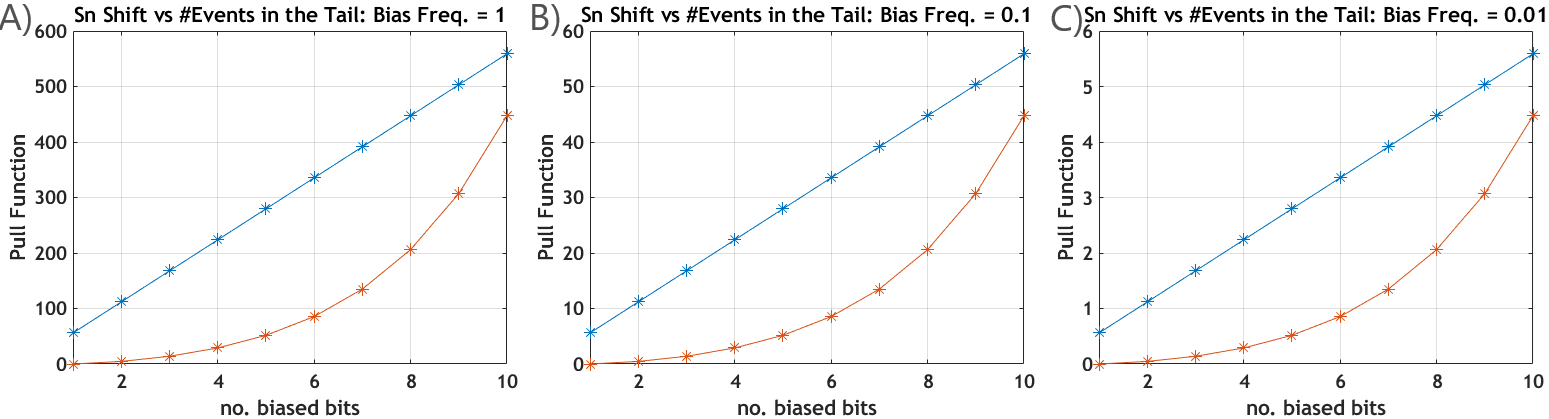}}
\caption{Comparison between the sensitivity of the estimators in the detection of a bias for the Monobit Test: the shift of the mean value of $\overline{S_n}$ (in blue) is evaluated against the variation in the number of anomalies expected (in orange), calculated as the integral of the distribution of events in the tails of the distribution over the $3\sigma$ limit, when a number of bits is forced to 1. The comparison is performed with the bias being introduced in every sequence (A), once every 10 sequences (B), and once every 100 sequences (C). The value of the pull function with respect to the unbiased sequence for the two estimators consistently identifies the shift on the mean value of $S_n$ as the most sensitive.}
\label{figPullFunction}
\end{figure*}

The diagnostic power of the Monobit on single bit-strings can be assessed as the capability to detect anomalies in the number of ones $n_1$ in a sequence of $n$ random bits, as those result in an absolute value of $S_n$ in excess of $k\sigma_{S_n}$, where $k$ determines the confidence level. This condition can be written as:
\begin{align}
\begin{split}
    S_{n} \geq k\sigma &\implies (2n_{1} - n) \geq k\sigma \\
    &\implies n_{1} \geq \frac{1}{2}(k \sqrt{n} + n).
    \label{eqn1ineq}
\end{split}
\end{align}

If $j$ bits are forced to one, the alarm is triggered whenever the number of ones in the remaining $n-j$ random bit string $n_{1}^{n-j}$ is
\begin{equation}
n_{1}^{n-j} \geq \frac{1}{2}(k\sqrt{n} + n) - j.
\end{equation}
The rate at which this happens can be computed as the tail of the distribution of $n_{1}^{n-j}$ and represents the True Positive Probability (TPP) for Monobit fails; on the other hand, the False Positive Probability (FPP) is associated to the statistical distribution of $n_1$ for an unbiased string. The comparison between these two quantities over a sequence of $n=32$ bits at varying confidence levels (ranging from 0 to $4\sigma$) is illustrated in Fig. \ref{false_vs_true_pos_k_scan}. Notably, the sensitivity is close to 50\% unless the number of biased bits grows to a significant fraction, reducing its usefulness. On the other hand, the analysis of a series of $S_n$ values can lead to a procedure enhancing the TPP, thus providing a basis for anomaly detection with single-bit precision. 

This consideration can be extended to all the tests described in the previous section, shifting the focus from the assessment of a single string to the statistical analysis of a series of $N$ sequences of $n$ random bits. This approach is aimed for an on-line differentiation between systematic failures, which indicate a bias in the source of bits, and occasional failures caused by statistical fluctuations. 

The observable considered is $\overline{S_{n}}$, the average value of $S_n$ over $N$ sequences, which is expected to be Gaussian distributed because of the Central Limit Theorem. A bias model is introduced by setting $j$ bits to 1 and, as outlined in Fig. \ref{figZShift}, two potential bias indicators can be considered: a normalized shift of the average value and a variation in the fraction of events in the distribution's tails. Presuming every sequence to be biased, the dependence of the shift from the number $j$ of biased bits is linearly dependent on $j$, in fact:
\begin{align}  
\begin{split}
    S_n &= 2n_1 - n = 2(n_{1}^{n-j} + j) - (n^{n-j} + j) = \\
    &= (2n_{1}^{n-j} - n^{n-j}) + j 
\end{split}
\end{align}
and  $E[\overline{S_{n}}]  = j$, eventually scaled by the fraction of biased sequences in the series. On the other hand, the fraction of events in the tail is expected to be non-linearly dependent on $j$, since it is the integral of the normalized $\overline{S_{n}}$ distribution above the threshold value $k$.
The sensitivity of the two measures is presented in Fig. \ref{figPullFunction} for an exemplary sequence length $n=32$ as the number of biased bits changes. The biasing frequency ranges from every sequence to one in every hundred, across series of $N=2^5$ sequences. Results prove that the $\overline{S_n}$ shift estimator outperforms the method of counting events in the tails. Setting a $k=3$ confidence level, the effect of a single biased bit can be detected for tampering frequencies higher than one in 10 sequences, while at least five biased bits are required when one in 100 sequences are biased and sensitivity is limited for lower frequencies, unless a higher statistics is considered.\\

\frenchspacing
\begin{table}[bp]
    \caption{Expected ISN (i.e., $\overline{ISN}$) given $k\sigma$ CL in a Normal Distribution.}
    \setlength{\tabcolsep}{2pt}
    \begin{tabularx}{\columnwidth}{{p{40pt}p{120pt}p{80pt}}}
    \toprule
    \textbf{\textit{$k$}} & \textbf{\textit{2-tailed failure probability}} & \textbf{\textit{$\overline{ISN}$}} \\
    \midrule
        1 & 0.317 & 3.15 \\
        3 & $2.7 \times 10^{-3}$ & 370.4 \\
        5 & $5.7 \times 10^{-7}$  & $1.7 \times 10^6$ \\
        7 & $2.56 \times 10^{-12}$  & $3.9 \times 10^{11}$ \\
    \midrule
    \end{tabularx}
    \label{tabGaussianApprox}
\end{table}

The series of the $\overline{S_n}$ values is the base of the on-line assessment of the quality of the bit-stream. If the current $\overline{S_n}$ exceeds the $3\sigma$ threshold, a warning is raised and the following values are considered. Since the probability of having two consecutive warnings is approximately $10^{-6}$, while the probability of having three is $10^{-9}$, the likelihood of such occurrences is negligible unless a bias in the generation process is present. 

\begin{figure}[t]
\centerline{\includegraphics[width=0.5\textwidth]{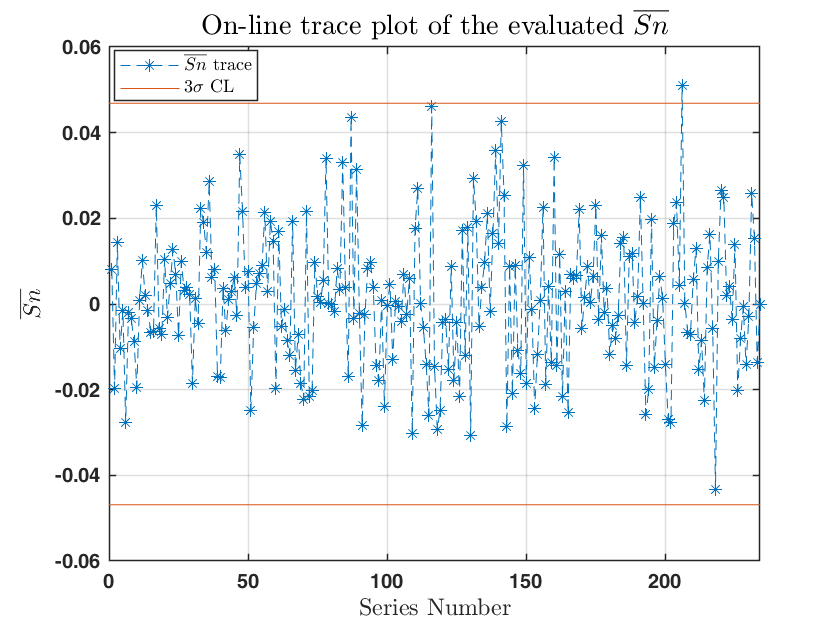}}
\caption{Trace plot of the computed $\overline{S_n}$ during on-line production over a number of series of $N = 2^{17}$ sequences of $n = 32$ unbiased bits for a total of 1Gb.}
\label{online_procedure}
\end{figure}

As exemplary illustration, 1Gb sample from the QRNG in use is partitioned into series, each with $N = 2^{17}$ sequences of $n = 32$ bits, for ease in computation in the FPGA implementation. In Fig. \ref{online_procedure}, the $\overline{S_n}$ trace plot during online production indicates no catastrophic failures of the entropy source, despite a single warning being triggered by statistical fluctuations. 

\begin{figure}[t]
\centerline{\includegraphics[width=.5\textwidth]{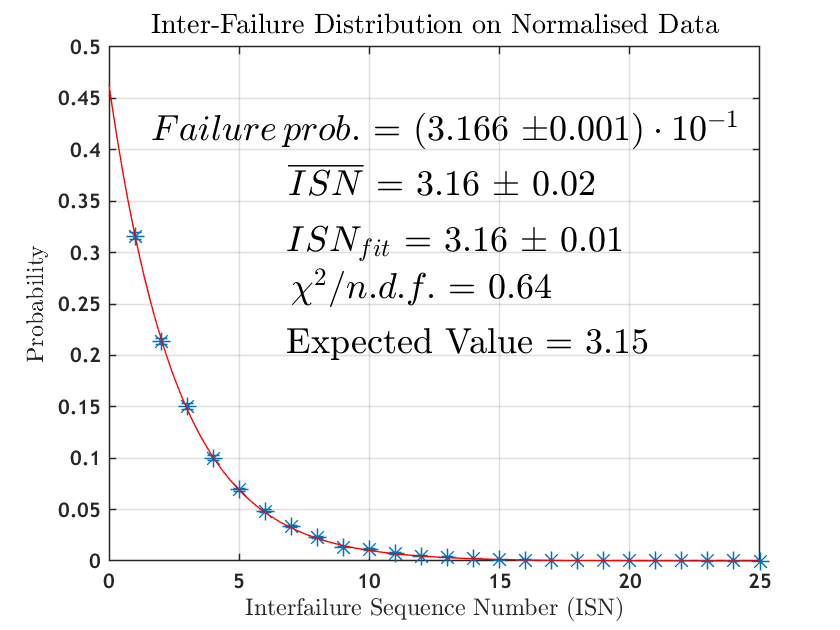}}
\caption{ISN trend of the Monobit test. Both the $\overline{ISN}$ and the computed $ISN_{fit}$ are compliant with the expected theoretical value for an unbiased sample.}
\label{figISNMono}
\end{figure}

Moreover, the bit quality was also assessed with a post-processing analysis of the collected data and measuring the distribution of the number of sequences in between two failures (ISN hereafter, for Inter-failure Sequence Number). The average value of ISN clearly depends on the threshold value over which a fail is declared, as reported in Table \ref{tabGaussianApprox}, so in order to have a large sample of failures the analysis is performed by lowering the $k$ value from 3 to 1. Results are shown in Fig. \ref{figISNMono}, where the trend is fitted with the model 
\begin{equation}
    P(X = x) = p  (1-p)^{x-1}, \label{eqMonoRunsFails}
\end{equation}
which describes the probability $P$ of having the next failure after $x$ sequences, $p$ being the failure probability. Results are statistically compliant with the hypothesis of an unbiased distribution, as confirmed by the goodness of the fit and the total number of failing sequences in 1Gb of data (measured to be $74.6 \pm 2.6$ against an expected value of 77).

\subsection{RUNS}

As for the Monobit, a collection of RUNS values was used to diagnose potential biases. Since, for each sequence, the expected average value depends on the actual number $n_1$ of bits set to 1, the measured number of runs $R_m$ is turned into the normalized variable provided by the z-score
\begin{equation}
    z = \frac{R_m - \overline{R}}{\sigma}, \label{eqdefZscore}
\end{equation}
which is expected to exhibit a Gaussian behaviour by the Central Limit Theorem. Unbiased series of $N$ sequences are expected to be centered around $\overline{z} = 0$.

Following the same approach as the Monobit test, the investigation of the sensitivity of the RUNS test is performed by introducing a bias in the number of runs in the sequences. 
By defining
\begin{align}
\begin{split}
    \overline{z} &= \frac{1}{N}\sum_{i=1}^{N} z_i,\\
    \sigma_{\overline{z}} &= \frac{1}{\sqrt{N}}, \label{eqAvgZscore}
\end{split}
\end{align}
presuming the sequence to be biased, an average change by $\Delta R$ in the number of runs will induce a variation
\begin{equation}
    \Delta \overline{z} = \frac{\Delta R}{\sigma_{\overline{z}}}, \label{eqDeltaZ}
\end{equation}
which can be identified as long as $| \Delta \overline{z} | \geq k \cdot \sigma_{\overline{z}}$ where $k$ is set according to the required confidence level. The threshold at $k$ standard deviations from the expected value of number of runs in a series is therefore
\begin{equation}
    | \Delta R | \geq k \cdot \frac{1}{\sqrt{N}}.
\end{equation}

The shift of the z-score due to the bias, with a frequency of once every 10 sequences, could be detected with bias at single-bit level with a confidence level of up to $5\sigma$. This level of sensitivity is maintained for sequences up to 128 bits long, while for longer sequences -- or equivalently for a lower frequency of the bias -- the sensitivity is limited and failure identification requires a larger deviation $\Delta R$ from the expected average.\\

\begin{figure}[t]
\centerline{\includegraphics[width=0.5\textwidth]{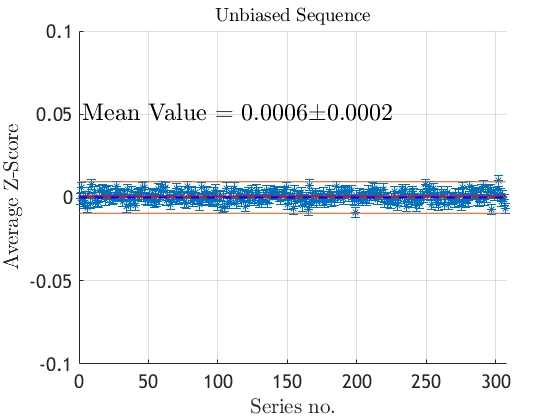}}
\caption{Trace plot of the z-scores with associated error bars computed during on-line production over 307 series of $N = 2^{17}$ sequences of $n = 32$ unbiased bits. Given the warning threshold of $3\sigma$, we expect 0.3\% of the population to generate warnings. The observed count of 1 warning aligns with this expectation.}
\label{runs_unbiased}
\end{figure}

\begin{figure}[tp]
\centerline{\includegraphics[width=0.5\textwidth]{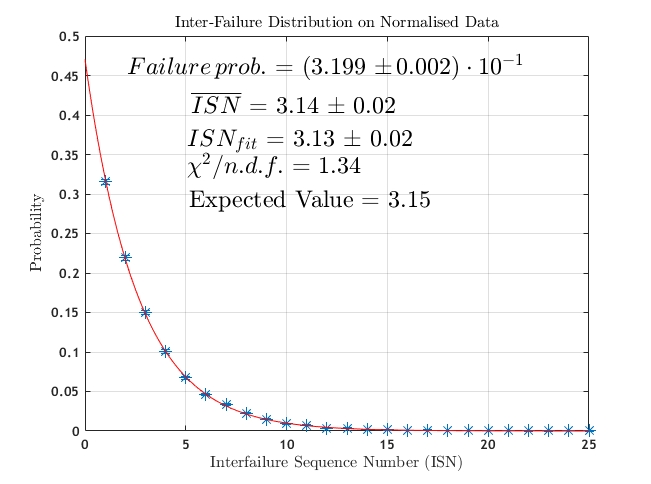}}
\caption{ISN trend for the RUNS test for different failure probability, fitted assuming $P(ISN = x) = p  (1-p)^{x-1}$. The resulting average $\overline{ISN}$ and fit parameter $ISN_{fit}$ yielded by the dataset, 1Gb of untampered random bits generated by the considered QRNG, are consistent with the expected value for an unbiased sample.}
\label{figResultsInterpolationRuns}
\end{figure}

Experimental results from the on-line evaluation of the z-score are displayed on the trace plot of Fig. \ref{runs_unbiased}. The computed z-scores remain within the expected confidence level, accounting for the statistical fluctuations. 
As for the Monobit, further analysis is performed by measuring the ISN. The results once more confirm the quality of the bit-stream, as reported in Fig. \ref{figResultsInterpolationRuns}.

\subsection{Repetition Count Test}

The Repetition Count Test is approached in a statistical framework by examining the properties of the distribution of the ISN, with the aim of assessing the quality of the bit-stream in terms of the min-entropy H. Due to the large amount of data necessary, this analysis is performed retrospectively.

Following NIST recommendation, the failure probability $\alpha$ is set to $2^{-20}$. Given the 4-bit symbols in the architecture, the threshold for consecutive identical symbols is $C=6$, as shown in Table \ref{tabCutThr}. Failures within a defined window are therefore expected to be driven by the Poisson distribution, with the interval between occurrences being exponentially distributed.

\frenchspacing
\begin{table}[b]
    \caption{Exemplary cut-off thresholds for multiple $H$ and $\alpha$ values, given $m=16$.}
    \setlength{\tabcolsep}{2pt}
    \begin{tabularx}{\columnwidth}{{p{80pt}p{80pt}p{80pt}}}
    \toprule
    \textbf{\textit{H}} & \textbf{\textit{$\alpha$}} & \textbf{\textit{C}} \\
    \midrule
        2 & $2^{-20}$ & 12 \\
        2 & $2^{-40}$ & 22 \\
        4 & $2^{-20}$ & 6  \\
        4 & $2^{-40}$ & 11 \\
    \midrule     
    \end{tabularx}
    \label{tabCutThr}
\end{table}

\begin{figure}[t]
\centerline{\includegraphics[width=0.5\textwidth]{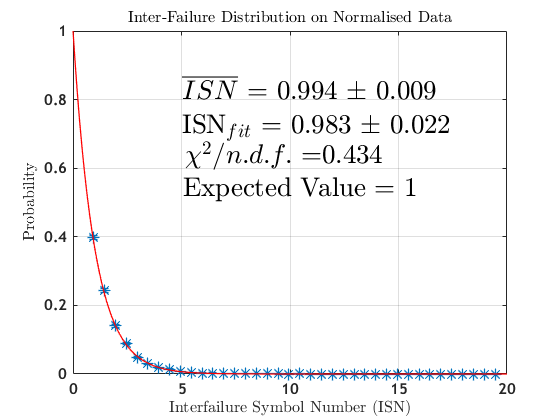}}
\caption{Distribution of the scaled $ISN$ for RCT on a 100Gb sample generated by one board of the QRNG under consideration, fitted with the exponential hypothesis. $ISN_{fit} \approx \overline{ISN}$ with 99.7\% confidence and both values are compatible with the expected scaled value of 1 for an unbiased bit-stream, corresponding to an average failure probability of $2^{-20}$.}
\label{figFitISN}
\end{figure}

Over 100Gb of random bits generated by four Silicon-based QRNG boards are evaluated, divided into 3 sets for each board. To avoid floating point approximation problems associated to exponentials with extremely low numbers, the measures were scaled w.r.t. the expected $ ISN = 1/\alpha = 2^{6\cdot 4-log_2(15)}$ obtained by applying \eqref{eqCRCT} with $C=6$ and $H=4$. As a matter of fact, the observed ISNs, normalized over the number of failures, conform with the exponential hypothesis, as shown by the fit in Fig. \ref{figFitISN}, with the computed parameter for the exponential distribution consistently falling within the 99.7\% confidence interval of the expected value for an unbiased source.

\begin{figure}[htp]
\centerline{\includegraphics[width=0.5\textwidth]{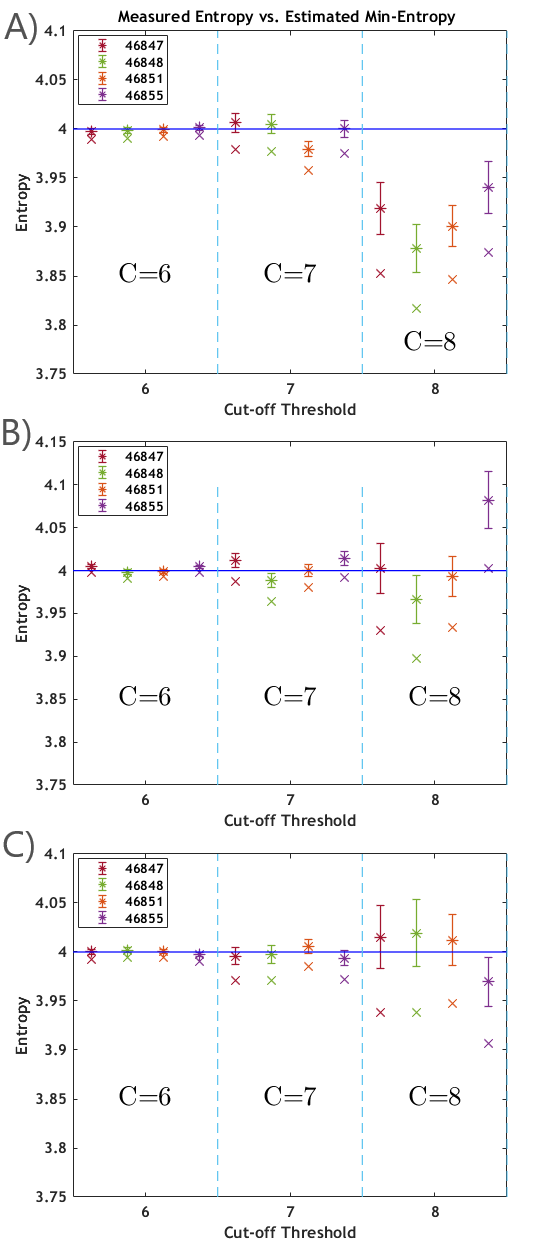}}
\caption{The plots exhibit the measured entropy, labeled as $\ast$, with the related uncertainty, and the min-entropy estimation, labeled as $\times$, which corresponds to lower bound on the measured entropy given a 99.7\% confidence, w.r.t. the physical limit $H = 4$. The values were computed on 100Gb of data produced with generators \#46847 (red), \#46848 (green), \#46851 (orange), and \#46855 (purple), divided into 3 sets (reported in Fig. \ref{Hshift} A, B, and C respectively), by running the test with 3 values of the cut-off threshold $C$. For $C = 6$, the min entropy estimations consistently fall between 3.989 and 3.998. Increasing the cut-off threshold diminishes the number of fails, thus increasing the error.}
\label{Hshift}
\end{figure}

The measured entropy $H$ and an estimation of the min-entropy are evaluated starting from $\overline{ISN}$. 
The measured entropy is obtained by replacing in \eqref{eqCRCT} the measured average failure rate $\overline{\alpha} = 1/\overline{ISN}$, as $C$ is fixed, and the uncertainty is computed by propagating the error:
\begin{equation}
    \sigma_H=\frac{1}{C\ln{2}}\frac{1}{\sigma_\alpha}.
\end{equation}
As the sample size of the number of failures is sufficiently large, the $\overline{ISN}$ is assumed to approximately follow a Gaussian distribution with $\sigma_{\overline{ISN}}=\sqrt{n_{fails}}$, where $n_{fails}$ denotes the total number of failures in the sample.
To determine a lower bound on the entropy, a shift of $3\sigma_{\overline{ISN}}$ to the right of the average is considered, in order to compute the largest possible true value of $\alpha$ (denoted as $\tilde{\alpha}$) compatible with a 99.7\% confidence to the observed one
\begin{equation}
    \frac{1}{\tilde{\alpha}} = \widetilde{ISN} = \overline{ISN} -3\sigma_{ISN}. \label{alphatilde}
\end{equation}
This value is then entered in \eqref{eqCRCT} to get the corresponding limit on the entropy. The results are reported in Fig. \ref{Hshift}, where different cut-off thresholds $C = 6, 7, 8$ are considered. Notably, the measured entropy does not consistently adhere to its physical constraints, reflecting the heuristic nature of the collected data, as shown in Table \ref{measured_h}. 

The min-entropy estimation represents the minimum entropy value that aligns with the measured outcome, computed by applying the aforementioned $3\sigma$ shift of the measured value. Taking $C = 6$, this lower bound consistently falls between 3.989 and 3.998 for each of the subsets of samples considered for all the boards. By increasing the cut-off threshold, the sample size diminishes, thus resulting in larger error bars. It is evident therefore that, by retaining the same confidence level, the lower bound decreases.

\subsection{Adaptive Proportion Test}

\begin{figure}[tp]
\centerline{\includegraphics[width=0.5\textwidth]{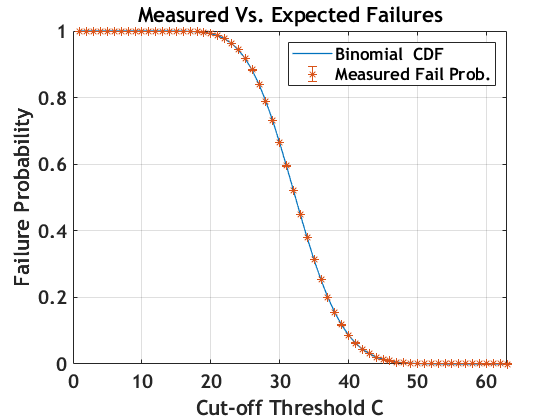}}
\caption{The performance of the APT over the unbiased bit-stream generated with Silicon-based QRNGs is investigated across a spectrum of cut-off thresholds. Notably, results consistently reveal that the measured failure probability remains bounded within the expected $3\sigma$ range of the Binomial CDF.}
\label{aptCDF}
\end{figure}

The analysis of the APT follows the same statistical approach of the RCT to estimate the min-entropy. As defined by NIST guidelines, the data is partitioned into sequences of $n = 512$ symbols. The number of occurrences of the first symbol within each sequence is counted, and the test is failed if the count exceeds a set threshold $C$. 
The failure probability therefore follows a Binomial Cumulative Distribution Function 
\begin{equation}\label{ATP_binoCDF}
	P(k \geq C) = B\_cdf\Big(C-1, N-1, p\Big),
\end{equation}
where $p$ is the probability of the selected symbol, assumed to be $\frac{1}{16}$ for an unbiased bit-stream. The test also considers cases where the frequency falls below a lower threshold. Given a failure probability $\alpha = 2^{-20}$ and $m=16$, the test fails if the count exceeds $C \geq 62$ or is below $C \leq 8$. The results, obtained for a varying the cut-off threshold, plotted in Fig. \ref{aptCDF}, align with the theoretical binomial cumulative distribution.

\begin{table*}[b]
  \centering
  \setlength{\tabcolsep}{3.9pt}
  \renewcommand{\arraystretch}{1.3}
\begin{tabular}{| *{10}{c|} }
    \hline
\multirow{ 2}{*}{Sets\textbackslash Boards}    & \multicolumn{2}{c|}{\#46847}
            & \multicolumn{2}{c|}{\#46848}
                    & \multicolumn{2}{c|}{\#46851}
                            & \multicolumn{2}{c|}{\#46855}                \\
    \cline{2-9}
  &   RCT  &   APT  &   RCT  &   APT  &   RCT  &   APT  &   RCT  &   APT  \\
    \hline
Set 1   &   $3.997 \pm 0.003 $  &   $4.007^{+0.014}_{-0.010}$  &   $3.999 \pm 0.003$  &   $3.990^{+0.009}_{-0.011}$  &   $3.999 \pm 0.002$  &   $4.004^{+0.010}_{-0.008}$  &   $4.001 \pm 0.003$  &   $3.955^{+0.011}_{-0.009}$  \\
    \hline
Set 2   &   $4.005 \pm 0.003$    &   $4.024^{+0.016}_{-0.011}$    &   $3.998 \pm 0.003$    &  $3.990^{+0.008}_{-0.009}$     &   $3.999 \pm 0.002$    &  $3.997^{+0.008}_{-0.007}$     &   $4.004 \pm 0.002$    &  $4.004^{+0.010}_{-0.008}$     \\
    \hline
Set 3   &  $4.000 \pm 0.003$     &  $4.006^{+0.012}_{-0.009}$     &   $4.002 \pm 0.003$    &  $3.994^{+0.011}_{-0.009}$     &  $4.000 \pm 0.002$     &   $3.992^{+0.008}_{-0.007}$    &   $3.997 \pm 0.002$    &   $4.012^{+0.012}_{-0.009}$    \\
    \hline
Weighted Avg.   &   $4.001 \pm 0.002$   &  $4.011 \pm 0.007$     &  $3.991 \pm 0.002$    &  $3.992 \pm 0.005$     &   $ 3.999 \pm 0.001 $   &   $3.997 \pm 0.004$    &  $ 4.006 \pm 0.002$    &   $4.003 \pm 0.005$    \\
    \hline
99.5\% CL   &   3.996   &  3.993     &  3.987    &   3.978   &   3.996   &   3.986    &  3.997   &   3.991    \\
    \hline
 min-H NIST   & \multicolumn{2}{c|}{3.999}
        & \multicolumn{2}{c|}{3.999}
            & \multicolumn{2}{c|}{3.998}
                    & \multicolumn{2}{c|}{3.999}                \\
\hline
\end{tabular}
\caption{A comprehensive report on the measured entropy for the RCT and APT for $\alpha = 2^{-20}$, showing that all values fall within the theoretical entropy limit, considering statistical fluctuations during the generation process, and assuming an unbiased bit-stream. The last row reports the min-entropy estimated according entropy source validation program by NIST \cite{nist800-90b}. The estimated values through the proposed method are lower due to the larger error.}
\label{measured_h}
\end{table*}

\begin{figure}[tp]
\centerline{\includegraphics[width=0.5\textwidth]{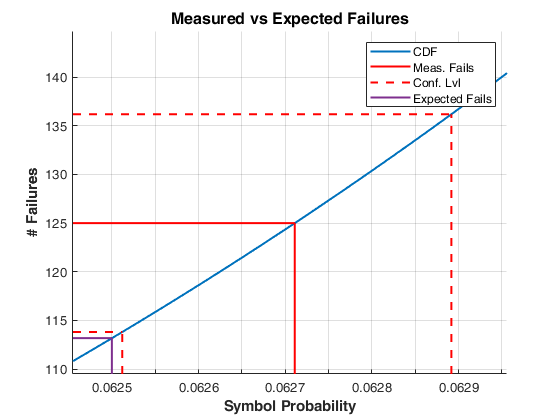}}
\caption{The estimation of $H$ is computed over a spectrum of probabilities $p$ given the expected and the measured number of failures on the bit-stream. The observed failures (alongside a one sigma binomial error) is traced back by the red line to the corresponding $p$, which yields $p = 0.0627\pm 0.0002$. The purple line marks the expected fails for the theoretical limit $H = 4$. Taking the +3 sigma value for the number of failures gives a lower bound for the entropy of the bit-stream of 3.998 at 99.7\% CL.}
\label{apt_p_max}
\end{figure}

The failure probability is linked to the probability of occurrence of a symbol $p$, which yields the bit-stream entropy as $H = -\log_2(p)$. Fig. \ref{apt_p_max} displays an exemplary value of estimated $p = 0.0627 \pm 0.0002$ based on measured failures and binomial error. The measured values for each set of each generator are reported in Table \ref{measured_h} for both the APT and the RCT. 

\section{Conclusions}
\label{sec:conc}

This paper reports the development of a statistical on-line implementation method for the Monobit and RUNS tests, which enables the detection of catastrophic failures of the entropy source with bias at single-bit level by monitoring shifts in the mean value during production. This method requires a relatively low amount of bits, as it was able to identify biasing on series of $2^{5}$ sequences of $32$ bits with a 99.7\% confidence level. The introduction of the ISN (Inter-Sequence Failure Number) and the analysis of its distribution are used for a retrospective analysis to verify the hypothesis.

The symbol-wise tests, based on the RCT and on the APT, are used to provide an estimate of the entropy of the bit stream. Four different boards of the QTRNG under test are evaluated, yielding consistent lower bounds for an unbiased bit-stream. 
As the NIST procedure follows the assessment of the IID properties of the bit-stream, it is computationally demanding. The proposed approach leverages on data already collected by implementing the NIST DRBG protocol, which provides an early-stage estimate of the lower bound of the entropy.

These advancements, along with the FPGA implementation of the tests, facilitate an efficient on-line evaluation of the bit-stream, ensuring it remains unbiased without compromising the bit generation rate. This comprehensive approach significantly enhances the reliability and efficiency of bit-stream generation in practical applications.

Future work could include evaluating the occurrences of each symbol in the alphabet for the APT, as it currently only takes into account the first symbol of the sequence. This is important as minimum entropy is linked to the occurrence rate of the most probable symbol. While the failure probability for each symbol is assumed to be binomial, correlations between symbols may affect failure counts, so the definition of precise thresholds is necessary and will be addressed in future research.

\section*{Acknowledgements}
This activity is part of the project named \textit{In-silico quantum generation of random bit streams (Random Power)} which has received funding from the European Union's Horizon 2020 Research and Innovation Programme within the ATTRACT cascade grant project, under the contract no.101004462).\\
The European Commission’ support does not constitute an endorsement of the contents, which only reflect the views of the authors. The Commission is not responsible for any use of the information therein.

\bibliographystyle{unsrt}
\bibliography{bibliography}

\end{document}